\documentclass[numberedappendix]{emulateapj}
\usepackage{natbib}
\usepackage{hyperref}
\hypersetup{colorlinks,citecolor=Blue,linkcolor=Red,urlcolor=Blue}
\usepackage{amsmath}
\usepackage{empheq}
\usepackage[usenames,dvipsnames]{color}

\allowdisplaybreaks

\begin{document}
 
\title{Feasibility of a resonance-based planet nine search}
\author{Elizabeth Bailey, Michael E. Brown, Konstantin Batygin}

\affil{Division of Geological and Planetary Sciences, California Institute of Technology, Pasadena, CA 91125}
\email{ebailey@gps.caltech.edu}

\newcommand{\Ham}{\mathcal{H}}
\newcommand{\G}{\mathcal{G}}
\newcommand{\appropto}{\mathrel{\vcenter{\offinterlineskip\halign{\hfil$##$\cr\propto\cr\noalign{\kern2pt}\sim\cr\noalign{\kern-2pt}}}}}
\newcommand{\Poincare}{{Poincar$\acute{\rm{e}}$}}

\begin{abstract}
It has been proposed that mean motion resonances (MMRs) between Planet Nine and distant objects of the scattered disk might inform the semimajor axis and instantaneous position of Planet Nine. Within the context of this hypothesis, the specific distribution of occupied MMRs largely determines the available constraints. Here we characterize the behavior of scattered Kuiper Belt objects arising in the presence of an eccentric Planet Nine ($e_9 \in 0.1$, $0.7$), focusing on relative sizes of populations occupying particular commensurabilities. Highlighting the challenge of predicting the exact MMR of a given object, we find that the majority of resonant test particles have period ratios with Planet Nine other than those of the form $P_9/P=N/1$, $N/2$ $(N \in \mathbb{Z}^+)$. Taking into account the updated prior distribution of MMRs outlined in this work, we find that the close spacing of high-order resonances, as well as chaotic transport, preclude resonance-based Planet Nine constraints from current observational data.
\end{abstract}
\maketitle
\section{Introduction} \label{sect1}

The primary line of evidence for the existence of ``Planet Nine,'' a distant, massive planet in the solar system, stems from the physical confinement of KBO orbits with semimajor axis in excess of $a \sim 250 \text{ AU}$ and perihelion distance beyond Neptune. While the dominant mode of dynamical coupling between Planet Nine and KBO orbits remains a subject of active investigation, it has been suggested that mean-motion resonances (MMRs) may widely occur in the distant, eccentric confined population of the Kuiper Belt. In particular, \cite{BB16a} were the first to recognize that synthetic particles exhibiting anti-alignment with Planet Nine exhibit temporary capture into MMRs with Planet Nine persisting over hundred-Myr time intervals.  Although \cite{Beust,Malhotra} have suggested that even non-resonant bodies might survive in a detached, anti-aligned state for the lifetime of the solar system, \cite{BatMorb} have found that, while secular effects are responsible for orbital clustering and perihelion detachment of KBOs in the confined population, MMRs are implicated in their long-term survival.

Explicit behavioral dependence on the period and phase of resonantly interacting bodies suggests that MMRs between Planet Nine and distant members of the scattered disk might encode information about the present-day semimajor axis and current position of Planet Nine, a possibility explored quantitatively by \cite{Millholland}. To this end, we note that the likelihood for a given object to occupy a specific resonance is fundamental to constraining the semimajor axis $a_9$ and subsequently the mean anomaly $M_9$ of Planet Nine. Therefore, one aim of this work is to produce a characterization of relative population sizes expected to occupy specific MMRs in the high-eccentricity case of Planet Nine.

While high-order mean-motion resonances\footnote{For $m/n$ resonance, the order is typically defined as $|m-n|$.} have negligible capture probability in the well-studied case of the circular restricted three-body problem, this tendency degrades in the high-eccentricity scenario of Planet Nine and, by extension, the population of KBOs sculpted by its gravity \citep{BB16a}. In other words, the typical ordering of terms found in the low-eccentricity expansion of the disturbing function \citep{MurraynDermott} breaks down for the problem at hand, warranting a numerical evaluation of capture probabilities for specific resonances. Finally, in light of the prior distribution of MMRs derived in this work, we explore the feasibility of updated resonant constraints on Planet Nine's present-day orbit and location.

The first potential resonance-based constraint is that on the semimajor axis of Planet Nine. Ideally, given the observed population of confined KBOs, it would be desirable to deduce likely period ratios with Planet Nine, thereby deriving the associated value of $a_9$. Using this brand of logic, \cite{Malhotra} reported an $a_9$ prediction of $\sim 665$ au, produced by supposing that the six confined objects known at the time reside in $P_{9}/P_{\text{KBO}}=N/1$ and $N/2$ period ratios with Planet Nine. Subsequently, \cite{Millholland} presented a more comprehensive analysis of the possibility of resonant constraints. Specifically, they constructed a $a_9$ distribution by assuming small integer ratio mean-motion resonances with $11$ known KBOs, followed by a Monte Carlo test to confirm the significance of the highest peak in their distribution. Enticingly, the highest peak in their $a_9$ distribution occurs around $660$ au, in apparent agreement with \cite{Malhotra}. The second potential resonance-based constraint on the present-day, instantaneous position of the planet in its orbit requires knowledge of the resonant angles themselves. Crucially, such a constraint would be highly valuable in the observational search for Planet Nine. 

This work is organized as follows. Section (\ref{sect2}) describes our suite of semi-averaged $n$-body simulations. Section (\ref{sect3}) describes the manner in which these simplified simulations capture the essence of the outer Kuiper belt's interactions with Planet Nine, and delineates the resulting numerically derived prior distribution of MMRs. The feasibility of resonance-based determination of Planet Nine's current location is further discussed in Section (\ref{sect6}). Concluding remarks are provided in Section (\ref{sect7}). 

\section{Two-Dimensional Numerical Simulations} \label{sect2}

\begin{figure}
\includegraphics[width=0.5\textwidth]{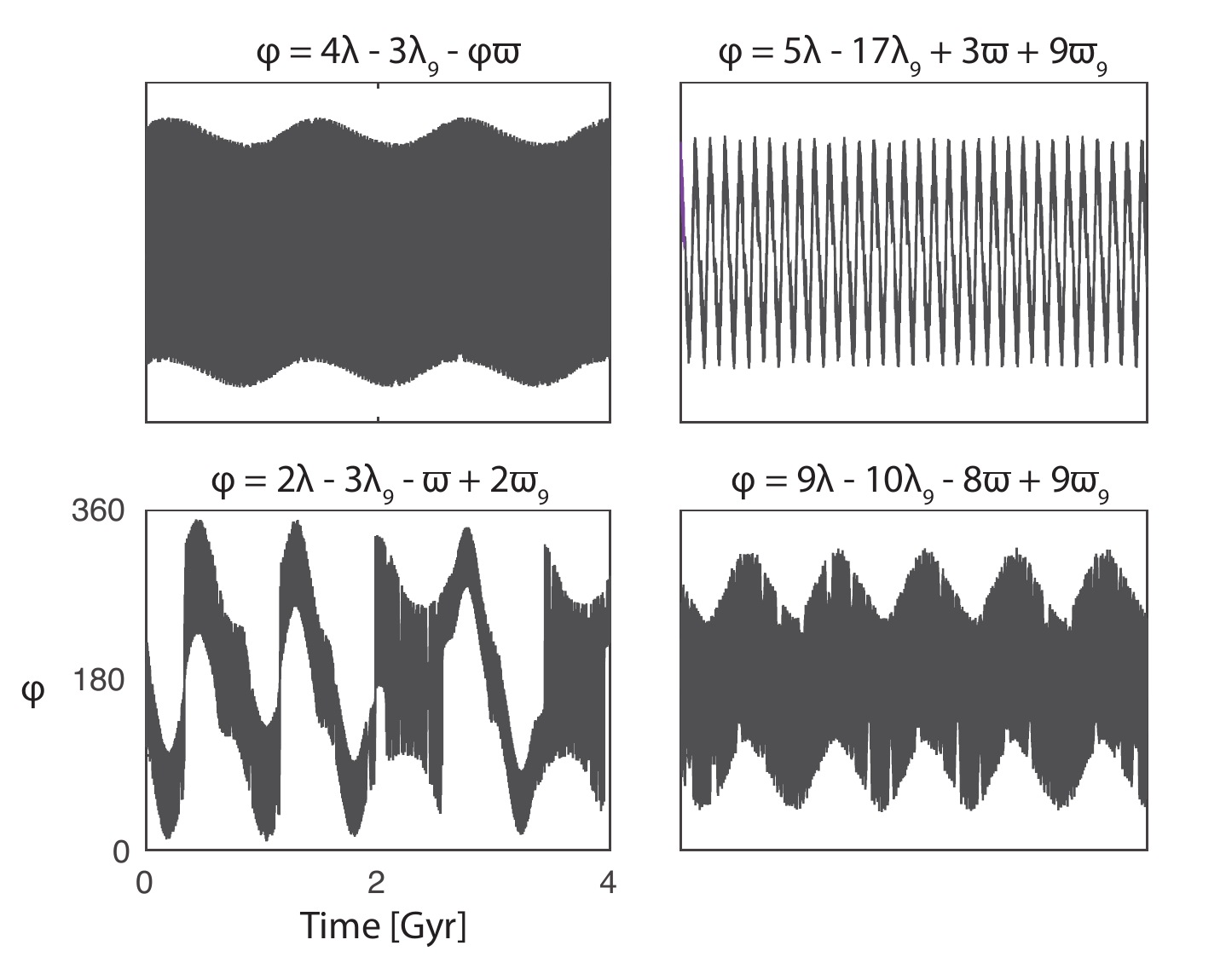}
\caption{Four examples of resonant angles $\varphi=j_1\lambda + F\lambda_9 + j_3\varpi + j_4\varpi_9$, for a variety of resonances.}
\label{resang}
\end{figure} 

\begin{figure}
\includegraphics[width=0.49\textwidth]{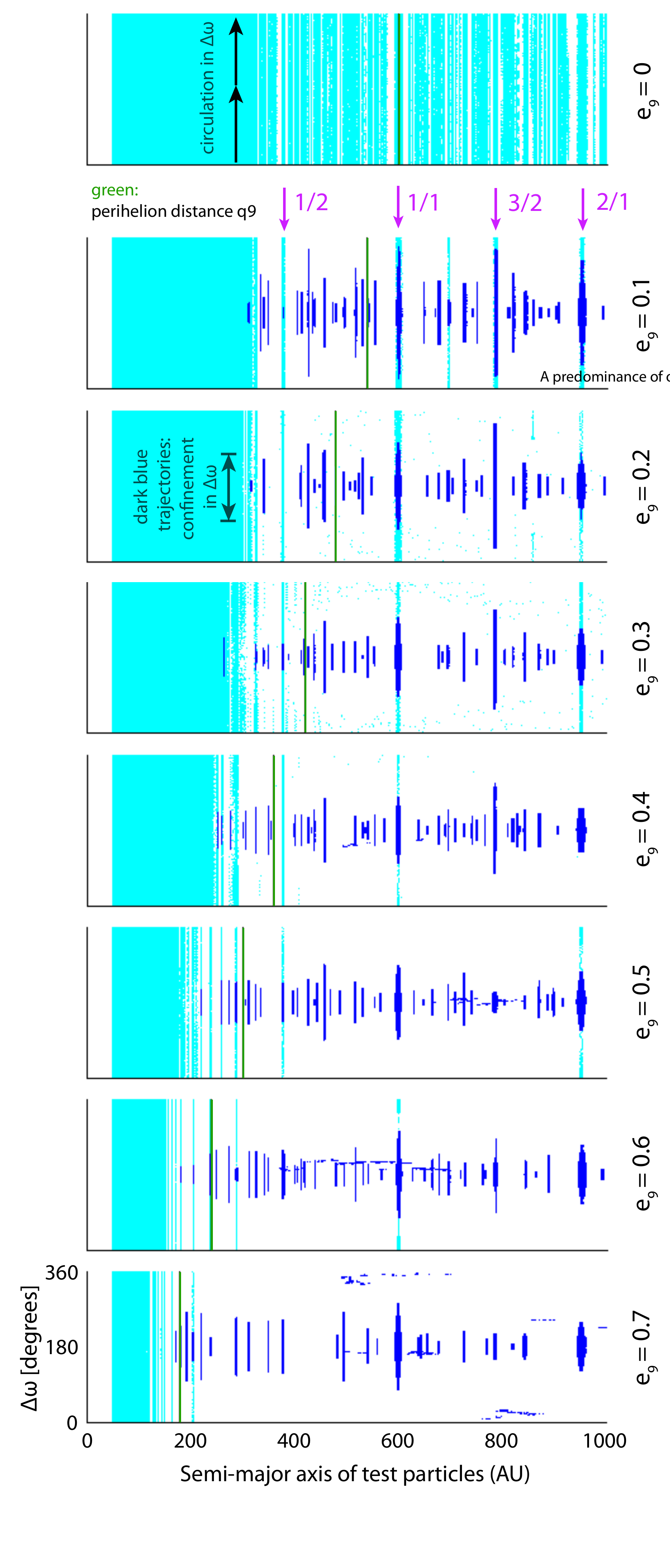}
\vspace{-4em}
\caption{Trajectories in semimajor axis and longitude of perihelion offset $\Delta \varpi$ for all bodies surviving the entire $4$-Gyr duration of simulations including a $10M_{\Earth}$ Planet Nine with $a_9=600$ au. The anti-aligned population (dark blue) is distinguished from other bodies (light blue) by libration in $\Delta \varpi$. Furthermore, the approximate radius below which confinement does not occur is typically lower than the perihelion distance $q_{9}$ of Planet Nine (green). Each plot corresponds to the result for a specific eccentricity $e_{9}$ of Planet Nine. Among simulations having an eccentric Planet Nine, several low-order resonances are preferentially occupied, including the $1/2$, $1/1$, $3/2$, and $2/1$ resonances. However, predominantly occupied are a variety of high-order resonances. }
\label{survbods}
\end{figure}

\begin{figure*}
\centering
\includegraphics[width=0.9\textwidth]{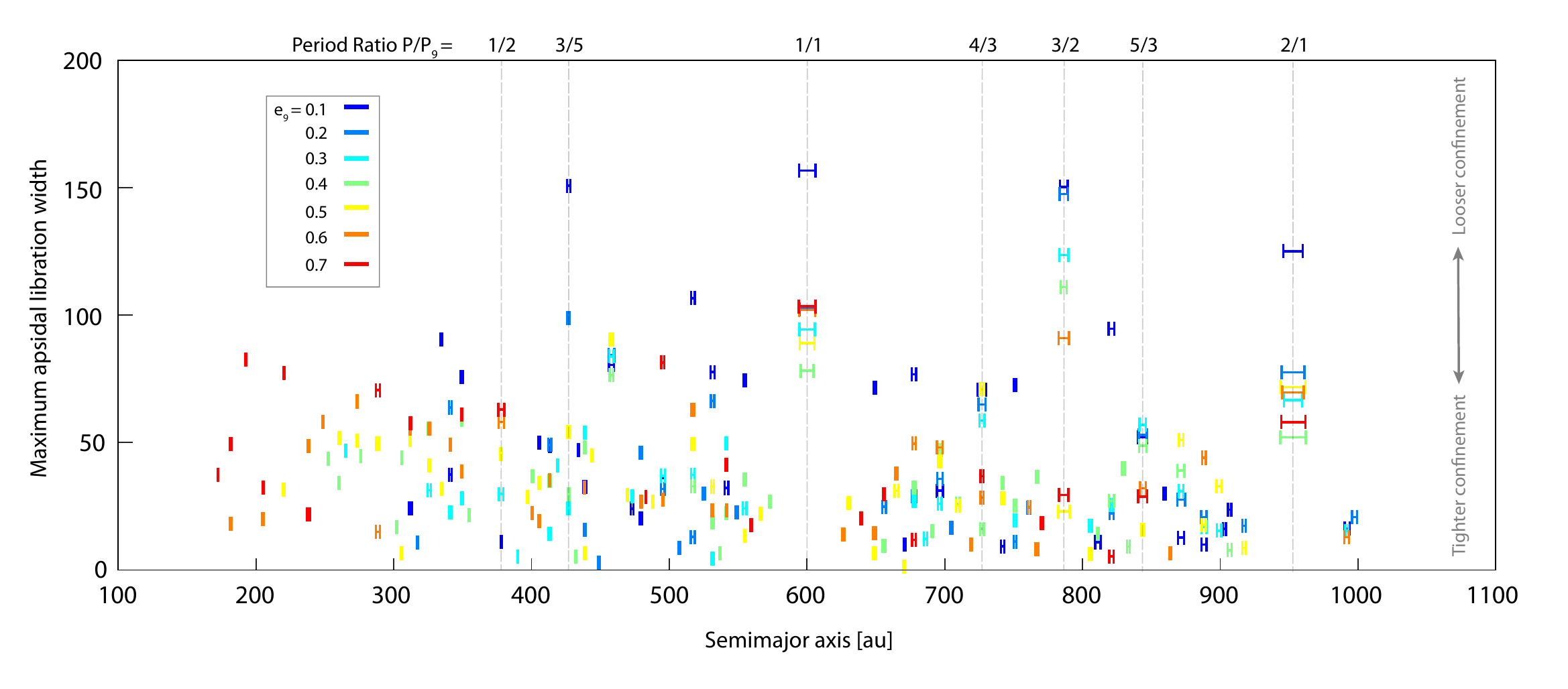}
\caption{The range of semimajor axis and maximum apsidal libration width exhibited in simulations by apsidally confined objects in specific resonances, across simulations featuring a range of Planet Nine eccentricities $e_9$.} 
\label{reswid}
\end{figure*}

By now, a considerable number of studies, employing variable levels of approximation aimed at simulating the dynamical evolution induced by Planet Nine, have been published in the literature \citep{BB16a, BB16c, BB16b, bailey, dlfm, Shank, Millholland, Lawler, Becker, BatMorb, Hadden}. It has been found, in $n$-body simulations accounting for the observed inclinations of distant KBOs and the $\sim 20-30$ degree inclination of Planet Nine, that objects tend to chaotically skip among commensurabilities. The primary aim of this work is to characterize a prior distribution of mean motion resonances in order to ascertain the feasibility of resonance-based constraints on Planet Nine, and such transitional behavior obfuscates the classification of specific MMRs. Thus, we employ a simplified, two-dimensional model of the solar system to understand the degree of resonance-based constraints that can be made. If significant resonance-based Planet Nine constraints can be obtained from present observational data, this capability should be best reflected in this highly idealized two-dimensional model.

Within the framework of this two-dimensional model, we confine all objects to the plane and average over the Keplerian motion of the known giant planets. Accordingly, the solar system interior to $30$ au is treated as a central mass with a $J_2$ gravitational moment having magnitude equivalent to the mean-field contribution of the canonical giant planets to the secular evolution of exterior bodies \citep{Burns,BB16a}. Hence, Planet Nine is the only massive perturber in these simulations. This model omits various realistic details. Notably, modulations in the eccentricity and inclination of KBOs due to close-range interactions with Neptune, as well as dynamics induced by the mutual inclination of the KBOs with Planet Nine, are absent from our calculations. Crucially, however, due to the lack of repeated transitions between resonances induced among surviving objects, these simplified simulations allow rigorous identification of the resonances in which objects reside, and their capacity to reveal Planet Nine's parameters.

We implemented direct $n$-body simulations using the mercury6 integration package \citep{Chambers}, employing the built-in Hybrid symplectic/Bulirsch-Stoer integrator \citep{Jwis, Press}, with time step chosen to be $1/8$ the orbital period of Neptune. In the simulations, we evolve an initially axisymmetric disk of eccentric test particles having uniformly random angular distribution and perihelion distance and semi-major axis randomly drawn from the $q \in [30,50]$ au and $a \in [50,1000]$ au range, respectively. While the initial distribution of test particles does not reflect the complete evolution of KBOs into resonance with Planet Nine, it serves as a probe of relative strengths of resonances. For each of the eight values of $e_9$ tested, $6000$ such test particles were randomly initialized and simulated. Particles attaining radial distances $r<10,000\text{ au}$ or $r<30\text{ au}$ were removed.

In principle, the relative strengths of resonances are not expected to vary significantly with $a_9$, as the relative strengths of individual terms associated with specific resonances in the usual expansion of the disturbing function only depend on the semi-major axis ratio of the interacting bodies \citep{MurraynDermott}. Therefore, Planet Nine was assigned a single, nominal semimajor axis, $a_9=600\text{ au}$. However, we note that different semimajor axes of $a_9$ would, in reality, subject the innermost resonances to variable levels of secular coupling with the canonical giant planets, altering the resonant widths slightly. Still, $a_9=600\text{ au}$ is roughly in keeping with the semimajor axis predictions of \cite{Millholland, Malhotra}; thus, we choose this value of $a_9$. Moreover, eccentricities $e_9$ were tested ranging from $0$ to $0.7$ in increments of $0.1$. The simulations in this work span $4$ Gyr in approximate accordance with the solar system's lifetime.

Because this work addresses the distribution of closely-spaced, high-order mean motion resonances, which have finite width in semimajor axis, the period ratio alone is insufficient to confirm a specific resonance. Instead, we confirmed specific mean-motion resonances with Planet Nine among the surviving objects by searching for a librating resonant argument (Figure \ref{resang}). The general form of such a resonant argument can be stated as $\varphi = j_1\lambda +j_2\lambda_9 + j_3\varpi +j_4\varpi_9$, where the d'Alembert relation, following from rotational symmetry, restricts the integer coefficients $j$ to satisfy $\sum_{i=1}^{4} j_{i}= 0$ \citep{MurraynDermott}. (For the $2$-dimensional case, we adopt the standard convention that longitude of ascending node $\Omega = 0$, thus $\varpi = \omega$.) Based on the values found for coefficients $j_1$ and $j_2$, identification of a critical argument informs the individual resonance in which a particle resides. Specifically, for an object in $p/q$ resonance with Planet Nine, the resonant argument takes the form $\varphi = q\lambda - p\lambda_9 + j_3\varpi +j_4\varpi_9$. 

\begin{figure}
\includegraphics[width=0.45\textwidth]{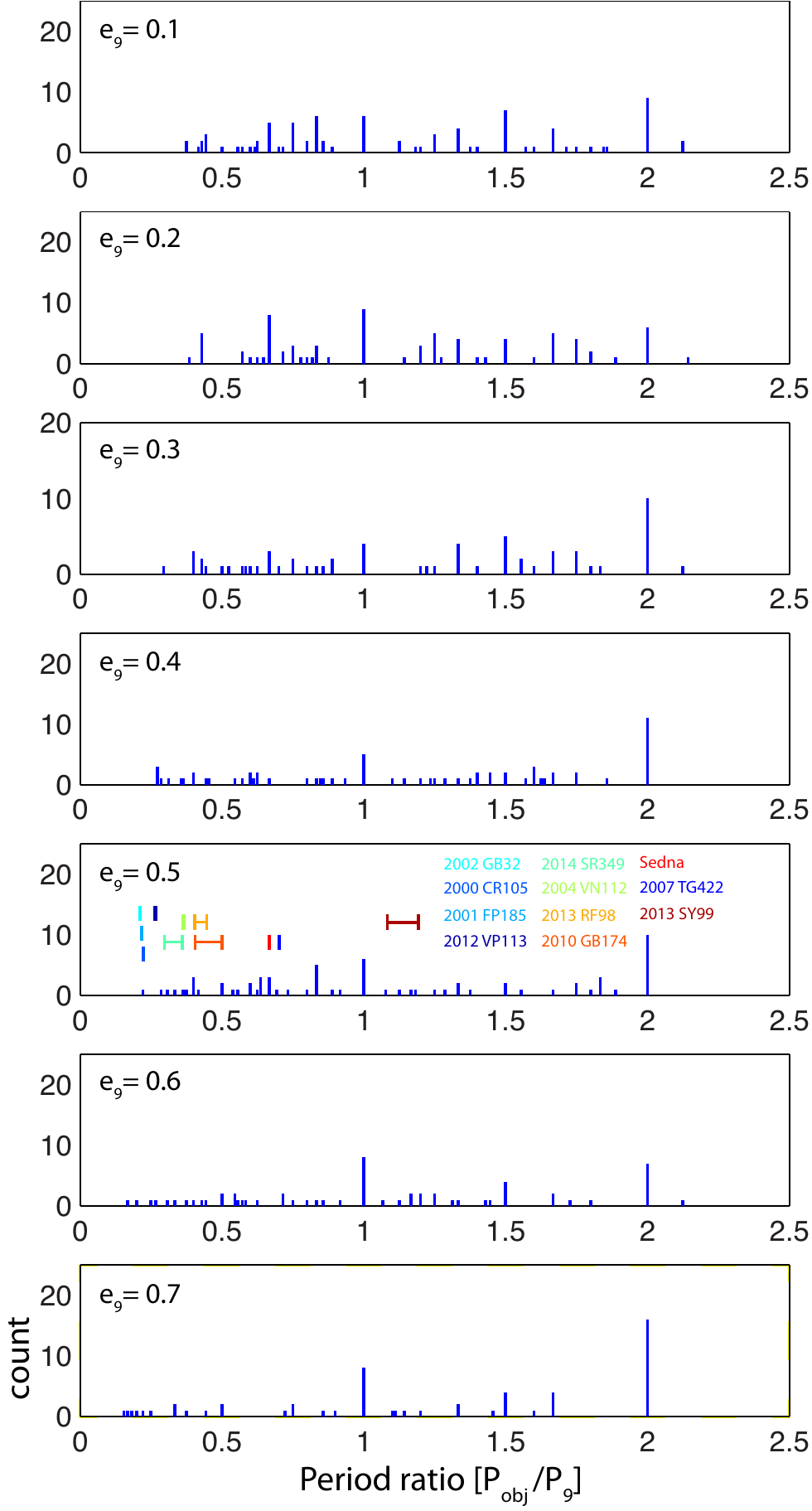}
\centering
\caption{Histogram with discrete bins showing the number of objects in each occupied resonance, for a range of Planet Nine eccentricities. Each bin is located at the exact commensurability ascertained by identification of resonant angles for objects. Note the close spacing of occupied high-order resonances. Beyond the axis bound, single objects at the $10/1$, $11/1$, $13/3$, $13/4$, $20/1$, and $22/7$ commensurabilities were also identified. The colored lines in the lower plot delineate the locations of commensurabilities predicted by \cite{Malhotra} (i.e. with Sedna at the interior $3/2$ resonance), with $1-\sigma$ observational error bars. Due to observational error in the KBO semimajor axes and the close spacing of occupied high-order resonances, we find there is no clear preference for this as opposed to many other resonant configurations.}
\label{histdenom}
\end{figure}

\begin{figure}
\includegraphics[width=.45\textwidth]{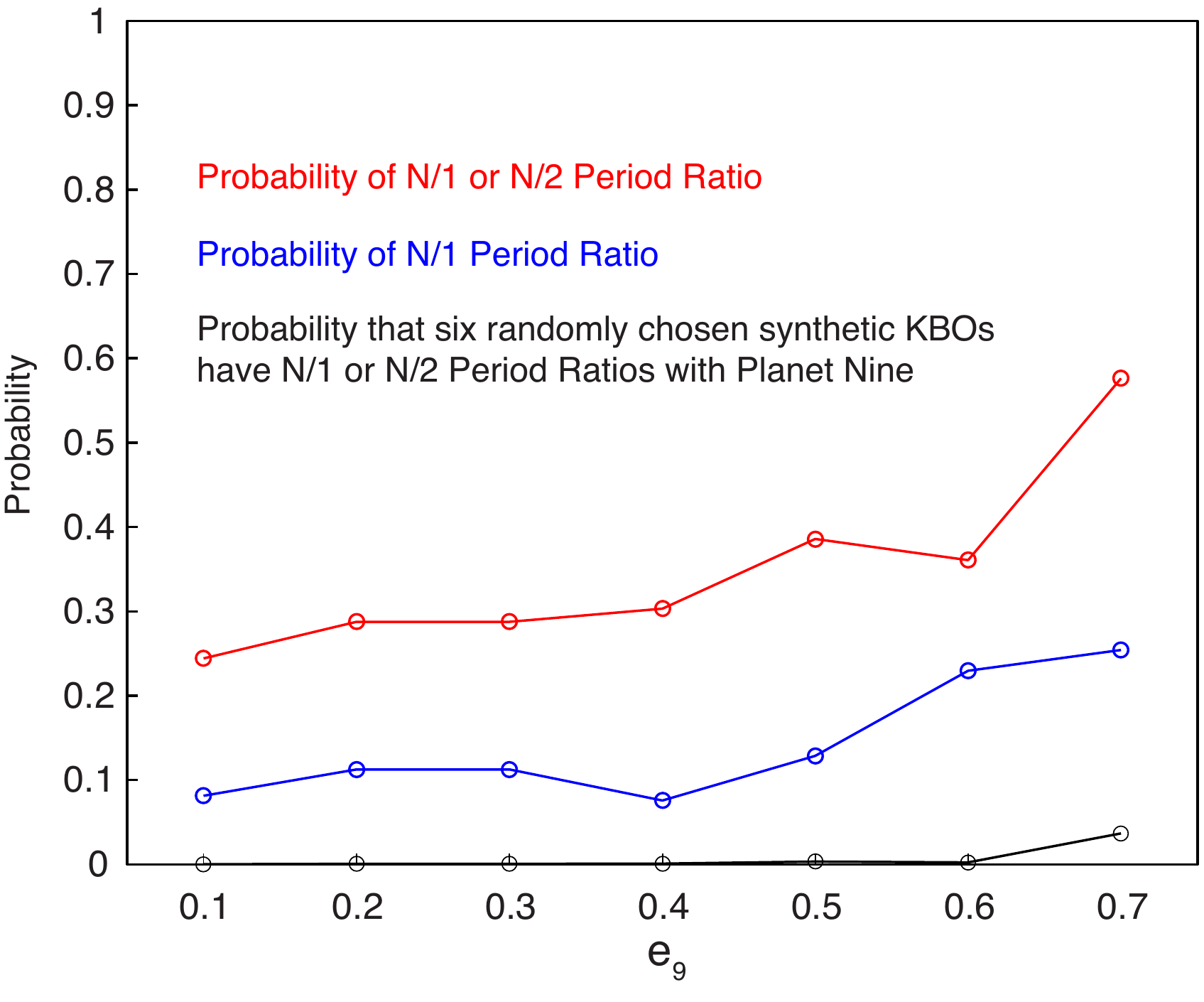}
\caption{Probability that a chosen synthetic particle has a period ratio $P_9/P = N/1\text{ } (N \in \mathbb{Z}^+)$ (blue), or a period ratio $N/1$ or $N/2$ (red). In particular, the probability that any six independently chosen objects will all have such period ratios is $\mathbb{P}(P_9/P \in \{N/1, N/2\})^{6}<0.05$, highlighting the prevalence of high-order resonances expected in the high-eccentricity case of Planet Nine.}
\label{NoverX}
\end{figure}

\begin{figure*}
\centering
\includegraphics[width=0.8\textwidth]{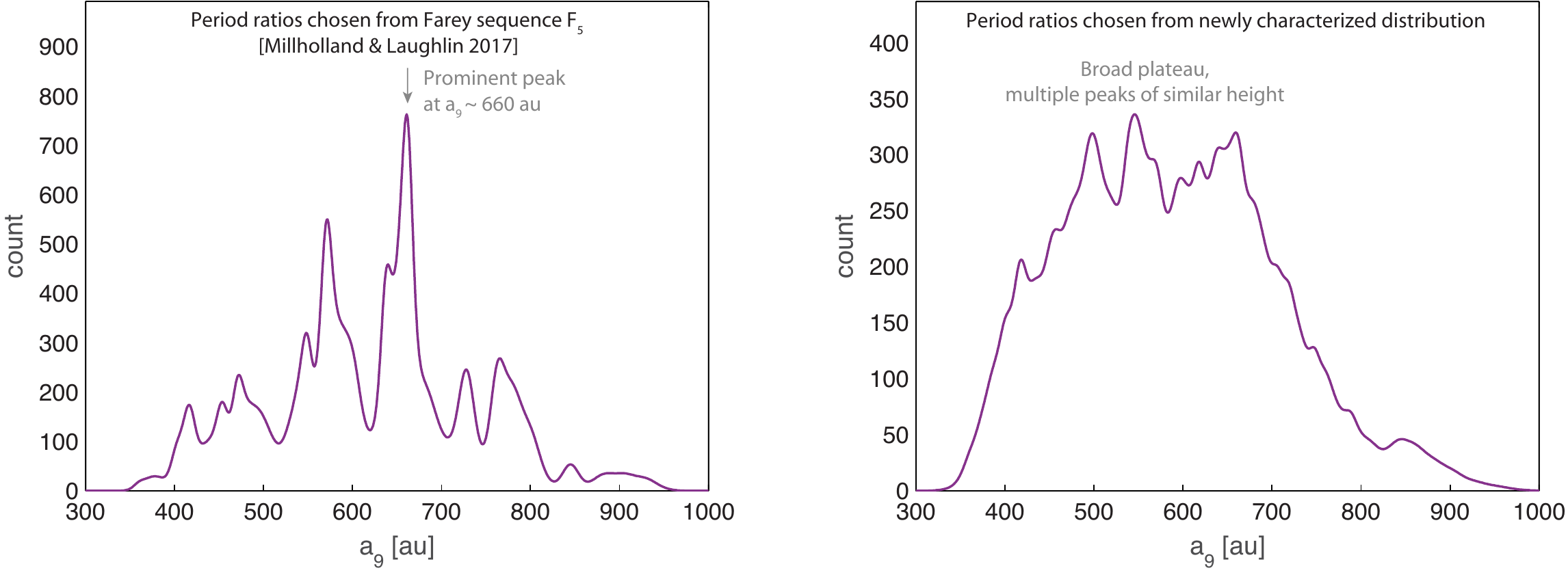}
\caption{Two distributions of the semimajor axis $a_9$, illustrating the difference invoked by considering the prior distribution of period ratios developed from the two-dimensional simulations in this work. \textit{Left:} Distribution developed by \cite{Millholland}. In constructing this distribution, the period ratios of observed objects were assumed to follow the distribution of the Farey sequence $F_5$ of period ratios having denominator $\leq 5$, with all such period ratios assumed equally likely. \textit{Right:} Distribution developed assuming the updated period ratio distribution. For details about the procedure invoked to produce these distributions, see \cite{Millholland}.}
\label{adistcomp}
\end{figure*}

\section{Behavioral regimes of surviving KBOs}
\label{sect3}

For the range $e_9\in[0,0.7]$ of Planet Nine eccentricities examined, the range of longitude of perihelion explored by test particles surviving for the full length of the $4$-Gyr simulations is shown in Figure (\ref{survbods}). For all cases of nonzero perturber eccentricity, the surviving bodies occupy several specific regimes. Below a critical semimajor axis $a_{\text{crit}}$, the vast majority of surviving bodies exist in a circulating regime, in which $\Delta \varpi$ covers all angular values. Notably, the value of $a_{\text{crit}}$ depends the semimajor axis, eccentricity, and mass of Planet Nine, and does not simply correspond to the perihelion distance of Planet Nine.

Objects with $a>a_{\text{crit}}$ exhibit orbital confinement, with the longitude of perihelion of the object relative to that of Planet Nine, $\Delta \varpi$, librating about $180$ degrees over the course of the simulation (i.e. in an anti-aligned configuration with respect to Planet Nine's orbit). This population is taken to be analogous to the clustered population of the observed distant Kuiper belt objects \citep{BB16a}, and will be the focus of the following sections of this paper.

While the majority of surviving bodies found in our two-dimensional simulations occupy the two aforementioned regimes, there are some exceptions. In particular, circulating objects are also found at distances significantly beyond $a_{\text{crit}}$--specifically, within a neighborhood of the $1/2$, $1/1$, $3/2$, and $2/1$ resonances with Planet Nine. Another class of objects observed is those that migrate stochastically through a variety of semimajor axes and values of $\Delta \varpi$ while avoiding ejection. Although fewer than $1$ in $600$ bodies exhibit this behavior in our two-dimensional simulations, such migratory evolution bears resemblance to the stochastic transport between commensurabilities observed among objects in full three-dimensional simulations.

The dynamical regimes occupied by test particles show broad-ranging consistency across all nonzero Planet Nine eccentricities examined. For example, Figure (\ref{reswid}) shows the maximum width in $a$ and widest apsidal excursion $|\omega-180^\circ|$ exhibited among objects in identified resonances. Notably, the width in $a$ is similar across the $e_9$ tested, and there appears not to be a strict relation between $e_9$ and confinement in $\Delta\varpi$. Instead, the primary indicators of $e_9$ in these simulations are the location of $a_{\text{crit}}$ and, in the two-dimensional case, the prevalence of circulating objects at the $1/2$, $1/1$, $3/2$, and $2/1$ resonances for lower-$e_9$ cases (Figure \ref{survbods}).

\subsection{Numerically derived period ratio distribution}

All confined objects in the simulation are in confirmed mean-motion resonance with Planet Nine. Moreover, the identified resonant angles persist over the full $4$-Gyr simulation. This stable behavior can be attributed to the lack of interference from close encounters with Neptune occurring in unaveraged three-dimensional simulations. Figure (\ref{histdenom}) shows the number of bodies found in individual resonances across the seven simulations with $e_9$ ranging from $0.1$ to $0.7$. In all simulations with $e_9>0$, the individual resonances with the greatest number of occupied objects were the $1/1$ and exterior $2/1$ resonances. However, as the next section will discuss, most objects do not occupy these particular commensurabilities. 

\section{Feasibility of resonant constraints on Planet Nine}
\label{sect6}

As Planet Nine's semimajor axis defines its period ratio with resonant KBOs, it is natural to attempt to predict $a_9$ through commensurabilities. Such predictions have been produced by \cite{Malhotra, Millholland}, in rough agreement at $a_{9} \sim 665$ and $a_{9} \sim 654$ au respectively. In particular, \cite{Malhotra} postulated that  $a_9 \sim 665$ au, placing Sedna at interior $3/2$ resonance and five other objects at $N/1$ and $N/2$ interior resonances with Planet Nine. However, the prior distribution of MMRs derived in this work appears to suggest that the majority of objects do not have period ratios with Planet Nine of the form $N/1$ or $N/2$ (Figure \ref{histdenom}). Furthermore, the probability that any six independently chosen objects all have $N/1$ or $N/2$ period ratios is less than $0.05$ (Figure \ref{NoverX}). We therefore conclude the assumption that distant KBOs are likely to reside in $N/1$ and $N/2$ resonances with Planet Nine is not supported by these numerical simulations.

We now turn to our numerically derived resonance distribution from our two-dimensional simulations and consider what information can be gathered about Planet Nine's whereabouts, following the statistical approach developed by \cite{Millholland}. Each iteration of their method proceeds as follows: First, they draw a sample $m_{i}$ from a truncated Gaussian mass distribution for Planet Nine, with mean $10 M_{\Earth}$ and bounds $5-20 M_{\Earth}$. Then, they randomly select one of the known distant KBO semimajor axes, together with one orbital period ratio from a chosen period ratio distribution. The implied Planet Nine semimajor axis is then calculated. If the implied semimajor axis lies in the range $[200 \text{ au} + 30m_i/M_\Earth, 600 \text{ au} +20m_i/M_\Earth]$ (derived based on \cite{BB16b}), a Gaussian centered at the implied semimajor axis, with $\sigma$ equal to the approximate resonance width, is added to the $a_9$ distribution being constructed. These steps are repeated until the resulting $a_9$ distribution is converged upon. In this way, \cite{Millholland} have provided a rigorous means of estimating $a_9$ by inferring mean motion resonances with the observed Kuiper belt objects. However, the period ratio distribution from which they drew their sample was simplified; they considered period ratios in the Farey sequence $F_{N}$, which includes all fractions with denominators up to $N$ (for example, $F_{4}=\{ \frac{1}{2}, \frac{1}{3}, \frac{1}{4}, \frac{2}{3},  \frac{3}{4}\}$). Critically, \cite{Millholland} made the assumption that period ratios in $F_{N}$ are equally occupied.  

In contrast, Figure (\ref{adistcomp}) illustrates the difference between the $a_9$ distribution reported by \cite{Millholland} using the Farey sequence $F_{5}$ distribution, versus the distribution obtained using the identical methodology and set of observed objects, but employing the period ratio distribution obtained in this work. We note that, because Planet Nine's eccentricity is poorly constrained, and because the period ratio distributions for varied $e_9$ resemble each other across the board, we have considered the sum distribution of period ratios found for all cases $e_9>0$. Due to the close spacing of many occupied high-order resonances, the prominent peak of the distribution of \cite{Millholland} is replaced in our $a_9$ distribution by a broad plateau. We note that reduction of peak prominence also occurs when the resonance distribution is taken to be an equally-weighted Farey sequence of higher order, although such treatment neglects the relative population occupying each resonance.

\begin{figure}
\includegraphics[width=0.46\textwidth]{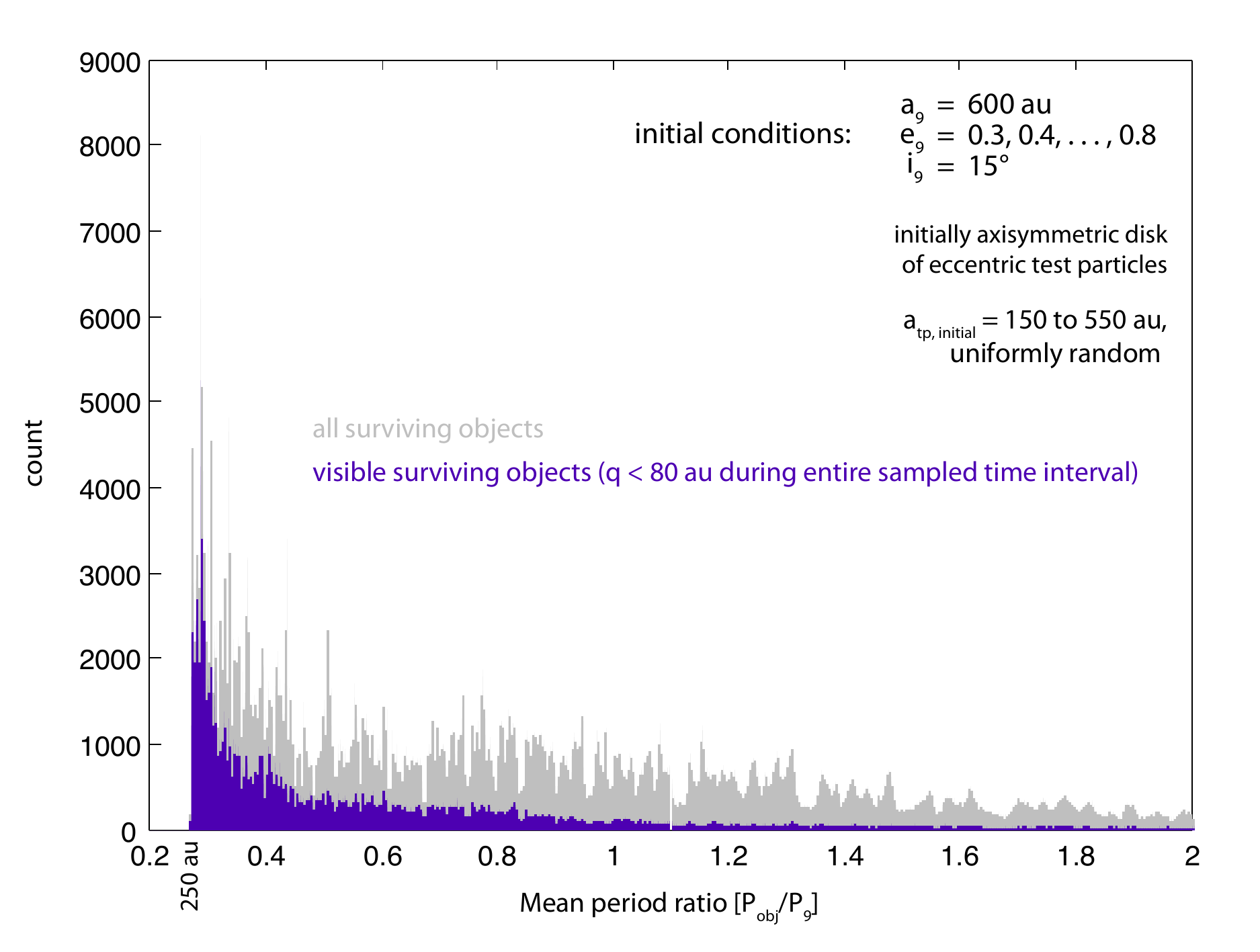}
\caption{Mean period ratio with Planet Nine of objects in full three-dimensional simulations, sampled in $1$-Myr intervals. Objects surviving the $4$ Gyr simulation, with perihelion distance $q>30$ and semimajor axis $a>250$ au, were considered. These simulations included all canonical giant planets of the solar system, in addition to Planet Nine. In order to avoid sampling the mean semimajor axis $\|a_{t.p.}\|$ of a test particle during scattering events, time intervals having $\text{max}(a_{t.p.})>\|a_{t.p.}\|+30 \text{ au}$ were excluded. A predominance of objects occupying any particular resonance is evidently lacking. This example suggests that the lessened predominance of low-order resonances in the high-eccentricity case of Planet Nine, demonstrated in the two-dimensional simulations of this work, continues to hold relevance in the realistic fully inclined case. }
\label{figgocompo}
\end{figure}

Indeed, our resulting plateau-shaped distribution for $a_9$ demonstrates that when the prevalence of high-order MMRs is considered, hopes of a resonance-based constraint on $a_9$ all but vanish, at least given current observations. Furthermore, compared to the simulations of a simplified Planet Nine system described in this work, additional behavior arises in fully inclined simulations which include the canonical giant planets (especially Neptune). Because the behavior exhibited by eccentric test particles in these more comprehensive simulations includes repeated transitions between resonances, fully rigorous determination of the distribution of objects among mean motion resonances is challenged. Moreover, a running average of period ratios with Planet Nine among objects in full simulations shows no obvious prevalence of objects occupying particular resonances (Figure \ref{figgocompo}). Without constraints on the period ratios between observed objects and Planet Nine, constraints on Planet Nine's present-day location along its orbit remain elusive.

\section{Conclusion}
\label{sect7}
Using two- and three-dimensional direct $n$-body simulations, we have highlighted several features inherent to the resonant dynamics of distant KBOs. Overall, we have illustrated that the resonance-based search for Planet Nine is frustrated by the increased strength of high-order resonances that follows from the high eccentricity of Planet Nine and the test particles. Existing work aiming to predict $a_9$ by considering possible period ratios between Planet Nine and observed objects \citep{Malhotra, Millholland} unfortunately neglects the prevalence of high-order resonances and chaotic transport--traits of the system which, in practice, challenge characterization of the specific resonances occupied. 

A constraint on $a_9$, to a degree allowing inference of the specific resonances in which observed objects reside, appears crucial for constraining the true anomaly of Planet Nine through resonant means. The degeneracy between mass and semimajor axis in simulations further complicates such efforts. Moreover, because knowledge of the specific period ratios between objects and Planet Nine are necessary to rule out Planet Nine's instantaneous position from specific regions of its orbit, resonance-based constraints on the current orbital position of Planet Nine do not appear feasible at this time.

The prevalence of high-order resonances in our simulations, brought about by the high eccentricity of the system, serves as a reminder that care should be taken before assuming the dynamics of circular orbits will hold for an eccentric case, such as Planet Nine's interaction with the confined population. For example, the lack of orbital angle clustering found by \cite{Lawler} in $n$-body simulations can be explained in part as resulting from their condition that $e_9=0$. It has already been established \citep{BB16a,BB16b} that Planet Nine's orbit must be eccentric to produce the confinement of distant, eccentric Kuiper Belt objects. Although it follows from symmetry, the results of our direct $n$-body simulations provide additional illustration of this fact (Figure \ref{survbods}). Because confinement occurs at the lowest nonzero Planet Nine eccentricity tested, $e_9=0.1$, the lowest eccentricity necessary to produce confinement in the planar eccentric three-body system is a possible subject of interest for future work. 

In summary, the expected phase-protected mean-motion resonances between observed distant KBOs and Planet Nine offer a tantalizing connection to the semimajor axis and current position of Planet Nine. Due to the high eccentricities involved, the underlying resonant dynamics are fundamentally different from the circular case, and high-order resonances with Planet Nine appear to dominate in the anti-aligned population. Considering this and taking the updated MMR distribution into account, the obtainable constraints on $a_9$ appear far less useful than suggested by \cite{Malhotra, Millholland}. A resonance-based constraint on the mean anomaly $M_9$ appears to require a constraint on $a_9$, so that correct resonant angles of observed objects can be deduced and specific regions of the orbit excluded. Thus, in addition to its unprecedented nature among planets in the solar system, the large eccentricity of Planet Nine currently postpones precise prediction of its position.

\section{Acknowledgments} We wish to thank Fred Adams and Sarah Millholland for useful discussions. In addition, we thank Caltech High Performance Computing for managing the Beowulf cluster \textit{Fram} and providing technical assistance, as well as the anonymous reviewer, for a thorough and insightful review that has led to a substantial improvement of the manuscript.

\end{document}